\documentclass[aps,prl,twocolumn,superscriptaddress,groupedaddress]{revtex4}  
\usepackage{graphicx}  
\usepackage{dcolumn}   
\usepackage{bm}        
\usepackage{amssymb}   

\def\ba{\begin{eqnarray}}
\def\ea{\end{eqnarray}}
\def\beq{\begin{eqnarray}}
\def\eeq{\end{eqnarray}}

\def\mpl{M_{\rm Pl}}

\def\p{{\cal P}}

\def\K{{\cal K}}

\def\L*{{\cal L}_*}
\def\L{\mathcal{L}}
\def\({\left(}
\def\){\right)}
\def\ie{{\it i.e. }}
\def\nn{\nonumber}
\def\p{\partial}
\def\mn{_{\mu \nu}}
\def\stu{St\"uckelberg }
\def\p{\partial}

\def\<{\langle}
\def\>{\rangle}
\newcommand{\eqref}[1]{(\ref{#1})}

\begin{document}

\title{Resummation of Massive Gravity}
\author{Claudia de Rham}
\address{D\'epartment de Physique  Th\'eorique, Universit\'e
de  Gen\`eve, 24 Quai E. Ansermet, CH-1211  Gen\`eve}
\author{Gregory Gabadadze}
\address{Center for Cosmology and Particle Physics,
Department of Physics, New York University,
NY, 10003, USA}
\author{Andrew J. Tolley}
\address{Department of Physics, Case Western Reserve University, 10900 Euclid Ave, Cleveland, OH 44106, USA}
\date{\today}


\begin{abstract}

We construct  four-dimensional covariant non-linear
theories of massive gravity which are ghost-free in the decoupling
limit to all orders. These theories resum
explicitly all the nonlinear terms of an effective field
theory of massive gravity. We show that away from the
decoupling limit  the Hamiltonian constraint is maintained at least
up to and including quartic  order in non-linearities, hence,
excluding the possibility of the Boulware-Deser ghost up to this order.
We also show that the same remains true to all orders in a
similar toy-model.

\end{abstract}

\maketitle



{\em \bf Introduction:} Whether there exist a consistent extension of General Relativity
by a mass term is a  basic question of a classical field theory. A small graviton
mass  could also be of a significant physical interest,
notably for the cosmological constant problem.

A ghost-free linear theory of massive
spin-2 -- the Fierz-Pauli (FP) model \cite{FP} -- had been notoriously hard to generalize to  the
nonlinear level \cite{BD}: the Hamiltonian constraint gets lost in general and,
as a result, the sixth  degree of freedom -- the Boulware-Deser (BD) ghost --
emerges as a mode propagating on otherwise physically meaningful local
backgrounds ({\it e.g.}, on a background of a lump of matter). Part of this problem
can be  seen in the effective field theory (EFT) approach to massive gravity   \cite{AGS}
in the decoupling limit  \cite{AGS,Creminelli}. There, the problem  manifests
itself in the Lagrangian for the helicity-0 component of the massive graviton.
This Lagrangian  generically  contains nonlinear terms with more than  two time derivatives.
The latter give rise to the sixth degree of freedom on local
backgrounds, while in general, these terms  lead  to the loss of well-posedness of the Cauchy problem
for the helicity-0 field theory \cite{AGS,Creminelli}.

A step forward has been made recently in \cite{deRham:2010ik} where it was shown that:
(a) the coefficients of the EFT can be chosen
so that the decoupling limit Lagrangian is ghost-free; this involves
choosing the ``appropriate coefficients'' order-by-order,
and an algorithm was set for this procedure to an arbitrary order;
(b) once  the ``appropriate coefficients'' are chosen  in the effective Lagrangian,
in the decoupling limit only a few  terms  up to the quartic order survive,
all the higher order terms vanish  identically. Moreover, the surviving
terms are unique as their structure is fixed by symmetries \cite{deRham:2010ik,dRGHP}.

In the present work we build on the above two points,
and go far beyond them. In particular:
(1) We  construct Lagrangians that
{\it automatically} produce  the ``appropriate coefficients'' once expanded in powers of the
fields; these give  rise to theories  that are  ghost-free automatically
to  all orders in the decoupling limit.
(2) Using the obtained Lagrangians we study the issue of the BD ghost
away from the decoupling limit;  we show that the Hamiltonian constraint
is maintained at least up to and including quartic order, hence
excluding the possibility of the BD ghost up to this order. We also express the exact potential
for gravity in a simplified (1+1)-dimensional model  and show explicitly how
the constraint is preserved to all orders.

The present framework provides explicit resummation of the nonlinear terms in the
EFT Lagrangian of massive spin-2. Another way to resum these terms
is to use an auxiliary extra dimension \cite{GG,Claudia}.
The latter has so far been shown to give
the ghost-free  decoupling limit only up to the cubic order
\cite{cubic}.  In \cite{GG,Claudia} the
resummation is obtained via the second  order partial
non-linear differential equation.  The present approach achieves this
via an algebraic non-linear equation.



{\em \bf Formalism:}
Define the tensor $H\mn$ as the covariantization of the
metric perturbation,
$ g\mn=\eta\mn+h\mn=
H\mn+\eta_{ab}\partial_\mu \phi^a \partial_\nu \phi^b$, where the four \stu fields $\phi^a$
transform as scalars, and $\eta_{ab}=(-1,1,1,1)$, \cite{AGS}.
The helicity-0 mode $\pi$ of the graviton can be extracted by
expressing   $\phi^a= (x^a-\eta^{a\mu}\partial_\mu \pi)$, such that
\ba
H\mn=h\mn+2\Pi\mn-\eta^{\alpha\beta}\Pi_{\mu\alpha}\Pi_{\beta\nu},~~~\Pi\mn\equiv \p_\mu \p_\nu \pi.
\ea
We may therefore define the following quantity
\ba
\label{Kmn}
\K^\mu_\nu (g,H)&=&\delta^\mu_\nu -\sqrt{\delta^\mu_\nu - H^\mu_\nu}
=-\sum_{n=1}^{\infty}d_n ( H^n)^\mu_\nu\,,\\
{\rm with} && d_n=\frac{(2n)!}{(1-2n)(n!)^2 4^n}\,.
\ea
Here $H^\mu_\nu = g^{\mu\alpha}H_{\alpha\nu}$, and
$(H^n)^\mu_\nu=H^\mu_{\alpha_1}H^{\alpha_1}_{\alpha_2}\cdots H^{\alpha_{n-1}}_\nu$ denotes the product
of $n$ tensors $H^\alpha_\beta$.
Below, unless stated otherwise, all the contractions
are made using the metric $g_{\mu\nu}$. The
tensor $\K\mn= g_{\mu\alpha}\K^\alpha_\nu$ is defined in such a way that
\ba
\K\mn(g,H)\Big|_{h\mn=0}\equiv \Pi\mn\,.
\ea
We use the same notation as in  \cite{Creminelli}
where square brackets $[\ldots]$ represent the trace of a tensor contracted using the Minkowski
metric, {\it e.g.} $[\Pi]=\eta^{\mu\nu}\Pi_{\mu\nu}$ and $[\Pi^2]=\eta^{\alpha \beta}\eta^{\mu\nu}\Pi_{\alpha\mu}\Pi_{\beta\nu}$, while angle brackets $\langle \ldots \rangle$ represent the trace with respect to the physical metric $g\mn$, so that $\langle H \rangle = g^{\mu\nu}H_{\mu\nu}$ and $\langle H^2 \rangle = g^{\alpha \beta}g^{\mu\nu}H_{\alpha\mu}H_{\beta\nu}$.

We are first interested in the decoupling limit. For that, let us define the canonically
normalized variables, $\hat \pi=\Lambda_3^3 \pi$  with $\Lambda_3^3=m^2 \mpl$ and $\hat h\mn=\mpl h\mn$.
The limit  is then obtained by taking $\mpl \to \infty$ and $m\to 0$ while keeping
 $\hat \pi$, $\hat h\mn$,  and the scale $\Lambda_3$ fixed. First, we construct an explicit
example of a non-linear theory that bears no ghosts in the decoupling limit, and then
give a general formulation and show the absence of the BD ghost beyond the decoupling limit in quartic order.

{\em \bf Massive Gravity:} The consistency of the
Fierz-Pauli combination relies on the fact that the Lagrangian
\ba
\label{L2der}
\mathcal{L}^{(2)}_{\rm der}&=&[\Pi]^2-[\Pi^2]\,,
\ea
is a total derivative.
To ensure that no ghost appears in the decoupling limit, it is sufficient
to extend $\mathcal{L}^{(2)}_{\rm der}$ covariantly away from $h\mn=0$, \ie replace $[\Pi]$ and $[\Pi^2]$
by $\<\K\>$ and $\<\K^2\>$ respectively, so that the total Lagrangian reads as
\ba
\mathcal{L}=\frac{\mpl^2}{2}\sqrt{-g}\(R-\frac{m^2}{4}\, \mathcal{U}(g,H)\)\,,
\label{L2}
\ea
with the potential $\mathcal{U}$ expressed as an expansion in $H$ as
\ba
\label{U2}
\mathcal{U}(g,H)&=&-4\(\langle \K\rangle ^2-\langle \K^2\rangle\)\\
&=&-4\big(\sum_{n\ge1}d_n\langle H^n\rangle\big)^2
-8\sum_{n\ge 2}d_n\langle H^n\rangle\,.\nn
\ea
Expanding this expression to quintic order,
\ba
&&\hspace{-5pt}\mathcal{U}(g,H)=\(\<H^2\>-\<H\>^2\)-\frac{1}{2}\(\<H\>\<H^2\>-\<H^3\>\)\hspace{5pt}\\
&&\hspace{-5pt}-\frac{1}{16}\(\<H^2\>^2+4\<H\>\<H^3\>-5\<H^4\>\)\nn\\
&&\hspace{-5pt}-\frac{1}{32}\(2\<H^2\>\<H^3\>+5\<H\>\<H^4\>-7\<H^5\>\)+\cdots \nn \,,
\ea
we recover the decoupling limit presented in \cite{deRham:2010ik} with the special indices $c_3=d_5=f_7=0$.

Note that the Lagrangian (\ref {L2}) with (\ref {U2}) can be obtained
from the Lagrangian
\ba
\mathcal{L}_\lambda =\frac{\mpl^2}{2}\sqrt{-g}\(R- {m^2}(\K_{\mu\nu}^2 - \K^2)\)  \nonumber
\\
+ \sqrt{-g} \lambda^{\mu\nu}( g^{\alpha\beta}\K_{\mu\alpha}\K_{\beta\nu} -2 \K_{\mu\nu} +H_{\mu\nu}),
\label{lambda}
\ea
where $\K_{\mu\nu} $ is an independent tensor field that gets related to $H_{\mu\nu}$  as in
(\ref {Kmn})  due to  the constraint enforced by the Lagrange multiplier $\lambda^{\mu}_\nu$.
Note, the expression (\ref {Kmn})  can be rewritten as  $\K^\mu_\nu = \delta^\mu_\nu
- \sqrt{\partial^\mu \phi^a \partial_\nu \phi^b \eta_{ab}}$, that gives a
square root structure in the full Lagrangian.

{\em \bf Decoupling limit:}
It is straightforward to notice that the leading contribution to the decoupling limit
\ba
\sqrt{-g}\, \mathcal{U}(g,H)\Big|_{h\mn=0}&=&-4\((\Box  \pi)^2-(\p_\alpha \p_\beta  \pi)^2\),
\ea
is a total derivative. The resulting interaction Lagrangian in the decoupling limit is then given by
\cite{deRham:2010ik}
\ba
\label{Lint def}
\mathcal{L}_{\rm int}
&=& \hat h\mn \bar X^{\mu\nu}\,,
\ea
with
\ba
\label{Xdef}
\bar X^{\mu\nu}=
-\frac{\mpl^2m^2}{8}\frac{\delta}{\delta h\mn}\( \sqrt{-g}\, \mathcal{U}(g,H)\)\Big|_{h\mn=0}\,.
\ea
Using the relations
\ba
\frac{\delta \K(g,H)}{\delta h\mn}&=&\frac12 \(g^{\mu\nu}-\K^{\mu\nu}\), \\
\frac{\delta \<\K(g,H)^2\>}{\delta h\mn}&=&H^{\mu\nu}-\K^{\mu\nu}\,,
\ea
the expression for $\bar X$ simplifies to
\ba
\bar X\mn&=&{1\over 2}
\Lambda_3^3\Big[\Pi\eta\mn-\Pi\mn+\Pi\mn^2 -\Pi \Pi\mn\\
&+&\frac 12 (\Pi^2-\Pi_{\alpha \beta}^2)\eta\mn\Big]\nn \,.
\ea
The tensor $\bar X\mn$ is conserved and gives rise to at most second order
derivative terms in the equations of motion. This tensor can be expressed as the product of two epsilon
tensors appropriately contracted with powers of $\Pi_{\mu\nu}$ \cite{dRGHP}.
For the potential \eqref{U2}, the Lagrangian in the decoupling limit is then given by, see
Ref.~\cite{deRham:2010ik}
\ba
\label{L decoupling}
\mathcal{L}^{\rm lim}_{\Lambda_3}=-\frac 1 4 \hat h^{\mu\nu}(\hat{\mathcal{E}} \hat h)\mn
+\hat h\mn \bar X^{\mu\nu}\,,
\ea
and this result is exact (\ie no higher order corrections). Notice that this is also in agreement with the results of \cite{deRham:2010ik} up to quintic order, for the special case $c_3=d_5=f_7=0$, but we explicitly demonstrate here that this result remains valid to all orders.

\vspace{10pt}
{\em \bf General formulation:}
As mentioned in \cite{deRham:2010ik}, at each order in the expansion there exists a total derivative contribution
\ba
\label{Lder n}
\mathcal{L}_{\rm der}^{(n)}(\Pi)=-\sum_{m=1}^{n}(-1)^m\frac{(n-1)!}{(n-m)!}\,
[\Pi^{m}]\,\mathcal{L}^{(n-m)}_{\rm der}(\Pi)\,,
\ea
with $\mathcal{L}^{(0)}_{\rm der}(\Pi)=1$ and $\mathcal{L}^{(1)}_{\rm der}(\Pi)=[\Pi]$.
These total derivatives generalize the ``Fierz-Pauli" structure used previously to all orders.
More generally, the potential of any theory of massive gravity with no ghosts in the decoupling
limit can be expressed non-linearly as
\ba
\mathcal{U}(g,H)&=&-4\sum_{n\ge2} \alpha_n \, \mathcal{L}^{(n)}_{\rm der}(\K)\,,
\ea
where $[\Pi^m]$ in \eqref{Lder n} should be replaced by $\<\K^m\>$ and expressed in terms
of $g$ and $H$ using \eqref{Kmn}.

Here again this specific structure ensures that the leading contribution to the decoupling
limit is manifestly a total derivative by construction,
\ba
\sqrt{-g}\, \mathcal{U}(g,H)\Big|_{h\mn=0}=\text{total derivative}\,,
\ea
and the resulting interaction Lagrangian can be derived by noticing the general relation
\ba
\frac{\delta}{\delta h^{\mu\nu}} \<\K^n\>\Big|_{h\mn=0}
=\frac n2 \(\Pi\mn^{n-1}-\Pi\mn^{n}\)\,,
\ea
so that
\ba
&&\hspace{-20pt}\frac{\delta }{\delta h^{\mu\nu}}\(\sqrt{-g}\mathcal{L}_{\rm der}^{(n)}(\K)\)
\Big|_{h\mn=0}=\\
&&
\sum_{m=0}^n\frac{(-1)^mn!}{2(n-m)!}\(\Pi^m\mn-\Pi^{m-1}\mn\)\mathcal{L}_{\rm der}^{(n-m)}(\Pi)\,,\nn
\ea
using the notation $\Pi^{0}\mn=\eta\mn$ and $\Pi^{-1}\mn=0$.
The decoupling limit Lagrangian is then given by \eqref{L decoupling} with the same definition
\eqref{Xdef} for the tensor $X\mn$, giving here
\ba
\bar X\mn= {1\over 2} \Lambda_3^3\sum_{n\ge 2}\alpha_n\(X^{(n)}\mn+nX^{(n-1)}\mn\)\,,
\ea
with
\ba
X^{(n)}\mn=\sum_{m=0}^n(-1)^m\frac{n!}{2(n-m)!}\Pi^m\mn \mathcal{L}_{\rm der}^{(n-m)}(\Pi)\,.
\ea
This is in complete agreement with the results obtained up to quintic order for $\alpha_2=1$,
$\alpha_3=-2c_3$, $\alpha_4=-2^2 d_5$ and
$\alpha_5=-2^3 f_7$. However we emphasize that the results in this paper are now valid to all orders.
The special theory found in \cite{GG,Claudia} corresponds to the specific choices of coefficients
$\alpha_2=1$ and $\alpha_3=-1/2$, see Ref.~\cite{deRham:2010eu}.

Furthermore, at each order the tensors $X\mn^{(n)}$ are given by the recursive relation
\ba
X^{(n)}\mn=-n \Pi_\mu^{\ \alpha}X^{(n-1)}_{\alpha\nu}+\Pi^{\alpha\beta}X^{(n-1)}_{\alpha\beta}\eta\mn\,.
\ea
with $X^{(0)}\mn=1/2 \eta\mn$. So since $X^{(4)}\mn\equiv 0$ all these tensors vanish beyond the
quartic one, $X^{(n)}\mn\equiv 0$ for any $n\ge 4$, and the decoupling limit therefore stops at that
order, as previously implied in \cite{deRham:2010ik}.

\vspace{10pt}
{\em \bf Boulware-Deser ghost:}
The previous argument ensures the absence of ghost in the decoupling limit, but it is
feasible that the ghost reappears beyond the decoupling limit, and is simply suppressed
by a mass scale larger than $\Lambda_3$. Certain arguments have
hinted towards the existence of a BD ghost,  \cite{Creminelli}. We reanalyze
the arguments  here and show the absence of ghosts within the regime studied.
To compute the Hamiltonian, we fix unitary gauge for which $\pi=0$, such that
\ba
\<H^n\>=\sum_{\ell\ge 0}(-1)^{\ell}C^{\ell+n-1}_{\ell}[h^{\ell+n}],
\ea
where the $C^n_m$ are the Bernoulli coefficients. We also focus on the case where $\alpha_2=1$ and $\alpha_n=0$ for $n\ge3$.
In what follows, we work in terms of the ADM variables \cite{Wald:1984rg},
\ba
g^{00}=-N^{-2},\ \ g_{0i}=N_i, \ {\rm and}\ g_{ij}=\gamma_{ij}\,,
\ea
with the lapse $N=1+\delta N$, and the three-dimensional metric $\gamma_{ij}=\delta_{ij}+h_{ij}$.
In terms of these variables, the potential is then of the form
\ba
\sqrt{-g}\, \mathcal{U}&=&\mathcal{A}+\delta N \mathcal{B}
+N_iN_j\big[-2 \delta^{ij}+\mathcal{C}^{ij}\\
&&+\delta N(\delta^{ij}+\mathcal{D}^{ij})-\frac 12 \delta N^2 \delta^{ij}-\frac 18 \delta^{ij}N_k^2\big]\,,\nn
\ea
where $\mathcal{A},  \mathcal{B},  \mathcal{C}^{ij}$ and $ \mathcal{D}^{ij}$ are functions of $h_{ij}$, at least first order in perturbations, and $\mathcal{C}^{ij}+2 \mathcal{D}^{ij}=-\frac 12 h^{ij}+\mathcal{O}(h_{ij}^2)$, and in this section we raise and lower the space-like indices using $\delta_{ij}$.
Notice that this is completely consistent with the analysis performed in \cite{Creminelli}, and corresponds to
setting the coefficients in (43) of \cite{Creminelli}  to $A=B=D=E=0$, while $C=-1/2$.
We emphasize here that the presence of a term of the form $C N_i^2 N^2$ does not signal the presence of a ghost,
since any quadratic terms in the lapse disappear after integration over the shift as we prove in what follows. Indeed, in terms of  redefined shift $n_i$,
\ba
N_j=\(\delta^i_j+\frac 12 \delta N\delta^{i}_{j}-\frac 18 \delta N h^{i}_{j}\)n_i\equiv L^{i}_{j} n_i\,,
\ea
the Hamiltonian is of the form
\ba
\label{Ham0}
\mathcal{H}=\frac{\mpl^2}{2}\sqrt{\gamma}\(N R^0+N_j R^j\)+\frac{m^2\mpl^2}{8}\(\mathcal{A}+ \mathcal{B} \delta N\) \\
-\frac{m^2\mpl^2}{4}L^{ij}\(n_in_j-\frac 12 \mathcal{C}^k_i n_j n_k+\frac1{16}n_k^2 n_in_j\)\nn,
\ea
up to quartic order in the metric perturbations.
Then, it is straightforward  to check that the variation of the Hamiltonian
(\ref{Ham0})  w.r.t. the shift   $n_i$ gives an equation which  is independent of
$N$, and  serves to determine $n_j$. Moreover, the lapse
remains a Lagrange multiplier even after integration over the shift, hence giving rise
to a Hamiltonian constraint on the physical variables. Whether this constraint gives rise to a
secondary constraint, and whether the system should be quantized as a first- or second class system,
is a separate interesting question. The mere existence of the Hamiltonian constraint is sufficient to claim
the absence of the BD ghost to that order
\footnote{The approach of \cite{Slava}  is equivalent to the EFT approach of \cite{AGS},
as was shown in \cite{BM}. Hence, the claim of \cite{Alberte:2010qb} on
the presence of the BD ghost in the quartic  order, if correct, would contradict our results. However, what has really been
diagnosed in \cite{Alberte:2010qb} is the issue already raised in \cite{Creminelli}, which we
have just addressed. In particular,
the apparent ghost-like nonlinear terms identified in \cite{Alberte:2010qb}, to the extent they were presented
in   \cite{Alberte:2010qb}, are in fact removable at that order by a nonlinear field redefinition,
in complete consistency with our results above.  This will be discussed in more detail elsewhere.}, yet without breaking Lorentz invariance, \cite{Rubakov:2008nh}.

The Hamiltonian evaluated on the constraint surface is proportional to
$m^2$ and whether or not it is positive semi-definite is determined by the
explicit expressions for $\mathcal{A}, \mathcal{B}, \mathcal{C}^{ij}$ and   $\mathcal{D}^{ij}$.  Thus, in general
certain backgrounds could have slow tachyon-like instabilities, however, this
is a separate issue from that  of the BD ghost that we clarified above.

{\em \bf $\bf (1+1)$-d massive gravity:} Proving the absence of the BD ghost in complete generality beyond
the quartic order is a grand task, which we save for a separate study. However, we can analyze here a
similar issue  in a $(1+1)$-d toy-model, where
we consider the Hamiltonian
\ba
\mathcal{H}=\mpl^2\sqrt{\gamma}\left[N R^0+\gamma^{11}N_1 R_1+\frac{m^2}{4}N \mathcal{U}(g,H)\right]\,,
\ea
with  $R^0$  and $R_1$ arbitrary functions of the space-like metric $\gamma_{11}$ and its conjugate momentum,
and the potential $\mathcal{U}$ is given in \eqref{U2}. In $1+1$ dimensions, it is relatively easy to
check that the Hamiltonian then takes the exact form
\ba
\mathcal{H}&=&\mpl^2\sqrt{\gamma}\Big[
N R^0+\gamma^{11}N_1 R_1- 2m^2 N \Big]\\
&&-2m^2\(1-\sqrt{(\sqrt{\gamma}+N)^2-\gamma^{11}N_1^2}\)\nn,
\ea
and seemingly includes terms quadratic in the lapse when working at quartic order and beyond,
\ba
\mathcal{H}\sim \mathcal{H}_0+ \mathcal{H}_1 N +m^2 N_1^2 N^2 +\cdots \,.
\ea
By stopping the analysis at this point one would infer that the lapse no longer enforces  a constraint.
However, this should be determined after integrating the shift. In other words, in terms of the redefined shift $n_1$
\ba
N_1=n_1 \, \(\gamma_{11}+N \sqrt{\gamma}\)\,,
\ea
the Hamiltonian takes the much more pleasant form
\ba
\mathcal{H}&=&\sqrt{\gamma}N R^0-2m^2\(1+\sqrt{\gamma}N\)\\
&&+\(\sqrt{\gamma}+N\)\(n_1 R_1+2m^2 \sqrt{1-n_1^2}\)\nn\,,
\ea
which remains linear in the lapse, even after integration over the shift. It is again straightforward
to see that the lapse does enforce a constraint, and does so for an ``arbitrary background".

{\em \bf Outlook:}
We have given  a covariant non-linear realization of massive gravity in 4D which:  (1)
is automatically free of  ghosts  in the decoupling limit,
to all orders in  non-linearities; (2) keeps the lapse as a Lagrange multiplier away from the decoupling limit, at least up to
quartic order in non-linearities. These findings constitute  what we believe is
a very significant  step forward,  and strongly suggests  the existence of an entirely ghost-free classical theory
of massive gravity.  However,  to prove this statement in complete generality,
two important ingredients are yet missing: (a) proving that the lapse remains a Lagrange multiplier
to all orders;  (b)  checking whether the secondary  constraint is generated or not, and
whether the theory could be canonically quantized as a first or second class system.
For the consistency of the theory at the quantum loop level one would have to
establish the existence of a symmetry which protects this theory against quantum corrections
that could revive the ghost. These points will be explored in a further study.

\emph{Acknowledgements:} We would like to thank M. Berg, C. Deffayet, S. Dubovsky, F. Hassan, D. Pirtskhalava and R. Rosen
for useful discussions. CdR is funded by the SNF and the work of GG was supported by NSF grant PHY-0758032.
CdR thanks the CoPS group at Stockholm University for its hospitality during completion of this work.



\end{document}